# Observation of a first order phase transition in fluid iron at pressures of 3 to 5 GPa

V.N. Korobenko, A.D. Rakhel[*]

Joint Institute for High Temperatures, Izhorskaya 13, Bld. 2, Moscow 125412, Russia

**Abstract**

Direct measurements of resistivity and caloric equation of state have been performed for fluid iron at pressures of 2 to 12 GPa in a wide density range. We found that the isochoric temperature coefficient of resistivity becomes negative, and this is considered as an indication of the metal-to-nonmetal transition, when density decreased by a factor of 3 to 4 compared to the normal solid density. We detected also that isentropes plotted in the pressure – specific volume plane have well-defined kinks localized on a convex curve with a maximum at about 5 GPa. Such behavior of isentropes evidences about a first order phase transition with a critical pressure one order of magnitude higher than the predicted pressure of the liquid-vapor critical point. Arguments are presented that the observed phase transition is most likely the liquid-vapor transition rather than an extra first order transition in the fluid state. We show that the gaseous nonmetallic phase represents dense plasma in the 1-2-th state of ionization so that it is a plasma phase transition as well.

Fluid metals when expanded sufficiently undergo transition into a non-metallic state [1,2]. At low temperatures this transition coincides with the liquid-vapor phase transition, while at high temperatures and pressures the independent first order metal-to-nonmetal (MNM) transition with its own coexistence curve and the critical point was predicted [3]. To the best of our knowledge, to date, for fluid metals any jumps of physical quantities in the MNM transition range have not been detected. Since mechanism of such transition as well as its effect on the thermodynamic functions is being poorly understood, the question about the existence of the first order MNM transition is of fundamental importance. In this letter we report about a first order phase transition detected in the MNM transition region of fluid iron. This transition occurs at pressures of about one order of magnitude higher than the estimated liquid-vapor critical point pressure and at the internal energy values which are 2-3 times larger than the sublimation energy of iron. It should be noted that apart from the fundamental interest our results could be useful as well for understanding of the Z-pinch experiments [4] in which the stainless steel wire-arrays are evaporated by an intense electric current pulse.

To study the MNM transition in fluid iron, direct measurements of its resistivity and caloric equation of state have been carried out. We used the pulsed Joule heating technique [5, 6]. As a result, the following two functional dependencies have been determined: resistivity $\sigma^{-1}$ ($\sigma$ is conductivity), and pressure $P$ as functions of specific volume $V$ and specific internal energy $E$. The essence of the experimental technique is as follows. An iron foil strip is tightly encapsulated between two flat sapphire plates and heated by an electric current pulse of high density. The geometric dimensions of the experimental essambly and the heating current pulse are chosen so that the distributions of temperature and density in the sample remain fairly uniform while the sample undergoes a 5 – 10 fold thermal expansion under a pressure of 2 - 10 GPa. The conditions of homogeneous heating are discussed in detail elsewere [7, 8]. This experimental technique was already used for the measurements on fluid aluminum and provided the fairly accurate results [5,6]. The schematic of our experiments, as well as the diagnostics is shown in Fig. 1.

In present experiments, the foil strips of pure iron (99,9% Fe) of 30 μm thickness, 3 - 6 mm width and 10 mm length, together with a ruby plate (with a thickness of 380 μm, a width of 10 mm and length of 10 mm), were placed between two optically polished sapphire plates of 1.5 - 5 mm thickness. The ruby plate was used to measure pressure by recording the ruby luminescence lines shifts [5,6]. The experimental assembly was carefully glued so that the thickness of the apoxy layer between the sample and the plates did not exceed 3 μm. On the side of the ruby plate facing the sample, a multilayer dielectric mirror of about 2 μm thickness was deposited. This mirror reduced the thermal radiation of the sample, that is an interference, and increased the signal of the ruby luminescence. In addition, it allowed to monitor possible melting of the ruby surface by the measurement of its reflection coefficient. In our experiments for the time of measurements this coefficient practically did not change, and hence, the ruby plate surface was not melted. The laser beam was focused so that the diameter of the spot on the dielectric mirror was 0.15 mm and that on the oposite side of the plate was 0.35 mm. The effect of heating the ruby plate and some other sources of uncertainties in the pressure measurements are discussed in Refs. 5,6.

To heat the iron samples a capocitor bank discharge was used. In each experiment we measured the current through the sample $I(t)$, the voltage drop across it $U(t)$ and the pressure near the sample $P(t)$. The sample resistance $R_s$, and the specific Joule heat dissipated $q$ were calculated as follows:

$$R_S(t) = U_R(t)/I(t), \quad q(t) = \int_0^t I(t')U_R(t')dt'/M$$

where $U_R(t) = U(t) - L_f \, dI(t)/dt$ is the active voltage drop, $M$ is the sample mass, and $L_f$ is its inductance. From the measured dependence $P(t)$ the sample volume was determined by solving the inverse problem about the motion of flat piston (the interface between the sample and sapphire) in a medium whose equation of state is known and the pressure on the piston is a given function of time. The equation of state of sapphire within the range of uniaxial elastic deformation ($P < 12.5$ GPa) is known with an uncertainty < 1%. As the difference between the mechanical properties of sapphire and ruby can be neglected (due to the low concentration of chromium in ruby), the integration of the equations of motion can be performed with almost the same accuracy [5]. After the sample volume has been calculated the mechanical work performed by the sample on the sapphire plates was determined. The specific internal energy $E$ is the difference between the specific Joule $q$ and the work per unit mass. Resistivity was calculated by the formula: $\sigma^{-1} = R_S(t)D(t)H/L$, where $D(t)$ is the sample thickness, $H$ and $L$ are its width and length (assumed to be constant).

Figure 2 shows the measured dependence of resistivity of fluid iron on specific internal energy. The marks represent resistivity values at fixed values of the relative volume $\varphi = V/V_0$, where the specific volume of iron under normal conditions $V_0 = 0.127$ cm$^3$/g. Every dashed line shows dependence measured in a certain experiment. In particular, the dependence for experiment # 24, in which pressure monotonically increased from 5 GPa to 11 GPa and experiment # 27 with a pressure about 2 - 3 GPa are indicated. The pressure effect consists in the systematically lower resistivity values for experiment # 24 compared to those for experiment # 27 in which pressure was essentially lower. As it can be seen, our measurements agree well with the data [9], obtained in the range $\varphi < 1.42$ at $P = 0.2$ GPa. It should be pointed out that the uncertainty in the resistivity values is less than 10% and that in specific internal energy doesn't exceed 15%. From Fig. 2, we can see that slopes of the isochores, which with good accuracy can be approximated by straight lines, at certain density change their sign. This means that the isochoric temperature coefficient of resistivity $\zeta_V = \left(\partial \sigma^{-1}/\partial T\right)_V$ also changes sign, since $\left(\partial \sigma^{-1}/\partial T\right)_V = c_V \left(\partial \sigma^{-1}/\partial E\right)_V$, and the heat capacity $c_V = (\partial E/\partial T)_V > 0$. Hence, the coefficient $\zeta_V$ is definitely positive for the isochores $\varphi = 2$, $\varphi = 2.5$ and turns to zero within the range $\varphi = 3 - 4$. For $\varphi > 4$, $\zeta_V$ becomes negative and much larger in magnitude than in the range where it is positive. Such behavior evidences, as we believe, about transition of iron into a nonmetallic state in which the temperature dependence of resistivity has a form: $\sigma^{-1} \propto \exp(\Delta/2kT)$, where $\Delta$ is the mobility gap in the electron density of states, and $k$ is the Boltzmann constant.

This conclusion is consistent with the criterion [10], according to which the mean free path of the conduction electrons $l$ in the metallic state at the MNM transition threshold approaches a minimum value of about the mean interatomic distance $d$ (the Ioffe-Regel limit). We can estimate $l$ using the measured values of resistivity and density by means of the Drude formula: $\sigma = n_e e^2 \tau / m$, where $n_e$ is the conduction electrons number density, $e$ is the electron charge, $m$ is its mass, and $\tau$ is the momentum relaxation time for the electronic subsystem. For a strongly degenerate system $\tau = ml/(\hbar k_F)$, where $\hbar$ is Planck's constant, and $k_F = (3\pi^2 n_e)^{1/3}$ is the Fermi wave vector, and $n_e$ is determined by the number density of ions $n_i$ and the number of conduction electrons per atom $z$: $n_e = z n_i$ (for liquid iron $z \approx 1,21$ [11]). The mean interatomic distance $d$ in the liquid state can be estimated considering atoms as dense random packed hard spheres, so that $\pi d^3 n_i / 6 = 0.63$ [12]. As a result we found that $l$ approaches $d$ within the interval $\varphi = 3 - 4$. It should be noted that the classical criterion of metallization [13], $\frac{4}{3}\pi n \alpha = 1$, where $\alpha$ is the atomic polarizability and $n$ is the number density of atoms at the metallization point, gives at this point $\varphi \approx 3$ (for $\alpha = 8.4$ Å$^3$ [14]). These estimates are discussed in detail in Ref. 15.

Fig. 3 shows the dependence of pressure on relative volume along isentropes. To obtain an isentrope starting at an initial point $(P_1, V_1)$ the equation $P = -(\partial E(V,P)/\partial V)_S$ was integrated numerically. As can be seen, some of the isentropes have kinks. The red curve in Fig. 3 shows a smooth approximation to the kinks positions. It is clearly seen that the curve saturates at a pressure about 4.7 GPa. It is well known, that for the liquid-vapor transition isentropes in the $(P,V)$ plane have kinks on the equilibrium line [16]. Therefore, Fig. 3 evidences about a first order phase transition in fluid iron with a critical point at 4.5 – 5.0 GPa. The question arises, what is the nature of this transition? Since the pressure is almost an order of magnitude higher than the estimated pressure of the liquid-vapor critical point (0.8 – 1 GPa [17 – 19]), and the values of internal energy on the equilibrium line of the observed here phase transition reach 18 kJ/g, that is more than twice the sublimation energy of iron ($E_{sub}$ = 7.4 kJ/g [20]), this transition seems not to be the liquid-vapor transition. As it was shown above, at expansions $\varphi \geq 4$ fluid iron undergoes transition into a non-metallic state so that it looks likely the kinks in the isentropes are due to the first order MNM transition, predicted in Ref. 3. We show below that, on a closer analysis, this conclusion should be recognized questionable.

Let's assume that the observed phase transition is the liquid-vapor transition and try to find out, is it a contradiction? Observations show, that for the liquid-vapor transition the temperature dependence of the saturated vapor pressure $P_{st}(T)$ in coordinates *log(P)* versus *1/T* is almost a linear function up to the critical point. In Fig. 4 the saturated vapor pressure data [21] and a linear fit to these data are plotted in these coordinates. Extrapolation of the fit line to $P$ = 4.7 GPa gives a temperature of 10.3 kK. The dependence $P_{st}(T)$ obtained in Ref. 22 gives a close value of the critical temperature so that an average value $T_c$ = 9.5 kK. It should be noted, that a 10% error in the critical pressure for this procedure results to a 2% error in the critical temperature.

According to the rectilinear diameter rule, the sum of the liquid phase density $\rho_L$ and the gaseous phase density $\rho_V$ on the equilibrium line is a linear function of temperature. At sufficiently low temperatures when $\rho_L \gg \rho_V$ we can neglect $\rho_V$ in the sum and, therefore, obtain $\rho_L(T) = aT + b$, where $a$ and $b$ are constants. Usually these constants are fit to the data on the thermal expansion coefficient of the liquid phase. We shall not do that because the constant $a$ cannot be found with good precision [23]. Instead of that, we obtain such dependence $\rho_L(P)$ which merges smoothly with the portion of the equilibrium line obtained in present work. In so doing we made use the handbook values of the boiling temperature (3.145 kK [21]), the liquid phase density at the boiling point (6.1 g/cm$^3$ [9, 21]), the critical temperature 9.5 kK and the saturated vapor pressure dependence $P_{st}(T)$ of Refs. 21, 22. The found in this way dependence $\rho_L(P)$ is shown in Fig. 3. It corresponds to $\varphi_c$ = 5.5.

Thus, we obtained the following set of the critical point parameters: $P_c$ = 4.7±0.5 GPa, $T_c$ = 9.5±0.8 kK, $E_c$ = 19±3 kJ/g, $\varphi_c$ = 5.5±0.6. It should be noted that only the critical pressure has been measured here. The other parameters were estimated using some uncertain procedures though based on our measurement results we can state that $\varphi_c > 4$ and $E_c > 16$ kJ/g. Using the critical parameters we can calculate the critical compressibility factor $Z_c = AP_cV_c/(RT_c)$, where $A$ is the atomic weight, and $R$ is the universal gas constant. The obtained value $Z_c$ = 2.3 exceeds essentially the value 0.285 used in Ref. 17 and indicates that fluid iron is strongly ionized at the critical point so that a remarkable contribution to pressure give the ionized electrons (since pressure exceeds remarkably the ideal gas value $RT/AV$). This conclusion is consistent with the large internal energy values at the equilibrium line which exceed essentially the sublimation energy and approach the ionization energy of ions with the charge state $z = 1 - 2$. Thus, the gaseous nonmetallic phase near the critical point represents a dense plasma so that the observed here transition is a plasma phase transition as well. It should be noted also that the degeneracy factor $T_F/T$ ($T_F$ is the Fermi temperature) along the equilibrium line decreases from 25 at the boiling point to a value about 6 at $\varphi$ = 2.5.

Finally, we calculated the critical entropy $S_c$ and compared it with the entropy of vapor on the equilibrium line at the boiling temperature $S_{vb}$. In the case of the liquid-vapor phase transition $S_c < S_{vb}$. To estimate $S_c$ we integrated the relation $dS = (dE + PdV)/T$ along the equilibrium line obtained here (from the boiling point to the critical point) and got $S_c - S_{lb} \approx 2$ J/(g·K), where $S_{lb}$ is the entropy of liquid at the boiling point. Since $S_{vb} - S_{lb} = Q/T_b$, where $Q$ is the vaporization heat at the boiling temperature ($Q = 6.27$ kJ/g [21]), we find $S_{vb} - S_{lb} = 2$ J/(g·K). The uncertainty of such estimates may approach 20%, hence, the obtained result $S_c \approx S_{vb}$ doesn't contradict our assumption that we observed nothing else than a liquid-vapor transition.

In summary, at expansions $\varphi = 3 - 4$ fluid iron undergoes transition into a nonmetallic state. In the range $\varphi = 3 - 6$ we detected a first order phase transition with a critical pressure about 5 GPa. We think that this transition is most likely the liquid-vapor phase transition rather than an extra first order phase transition in the fluid state.

We should like to thank Dr. W. J. Nellis for constructive suggestions and comments. This work was supported by the Basic Research Program of the Presidium of RAS P-09 "Study of matter under extreme conditions", as well as the Russian Foundation for Basic Research, grant № 06-08-00304-a.

[*] Electronic address: rakhel@ihed.ras.ru

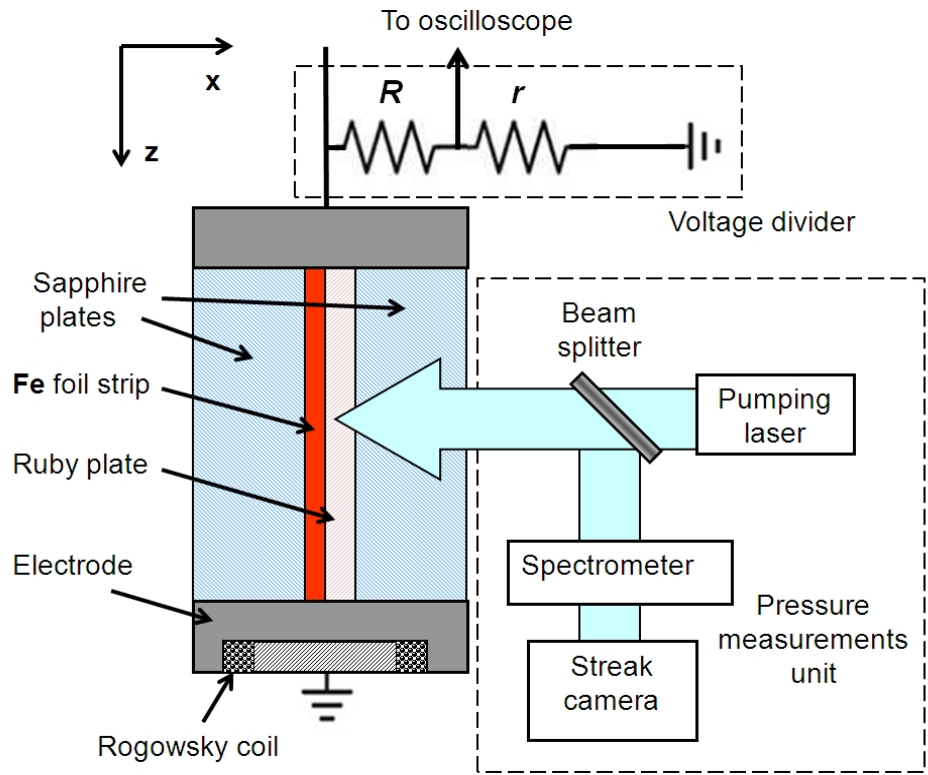

**Fig. 1.** (Color online) Schematic of the experiment. The experimental assembly (a flat iron foil strip sample placed together with a ruby plate between two sapphire plates) is shown in the plane perpendicular to the surface of the foil strip. Current flows along the *z* axis, and the sample undergoes thermal expansion mainly along the *x* axis because its thickness is much smaller than width and length and because it is sandwiched between flat sapphire plates. The current through the sample is measured with the Rogowsky coil and the voltage across the sample is measured with the voltage divider. The pressure in the sample is measured by recording the luminescence lines shifts of ruby compressed by the expanding sample. Luminescence in ruby is pumped by a laser.

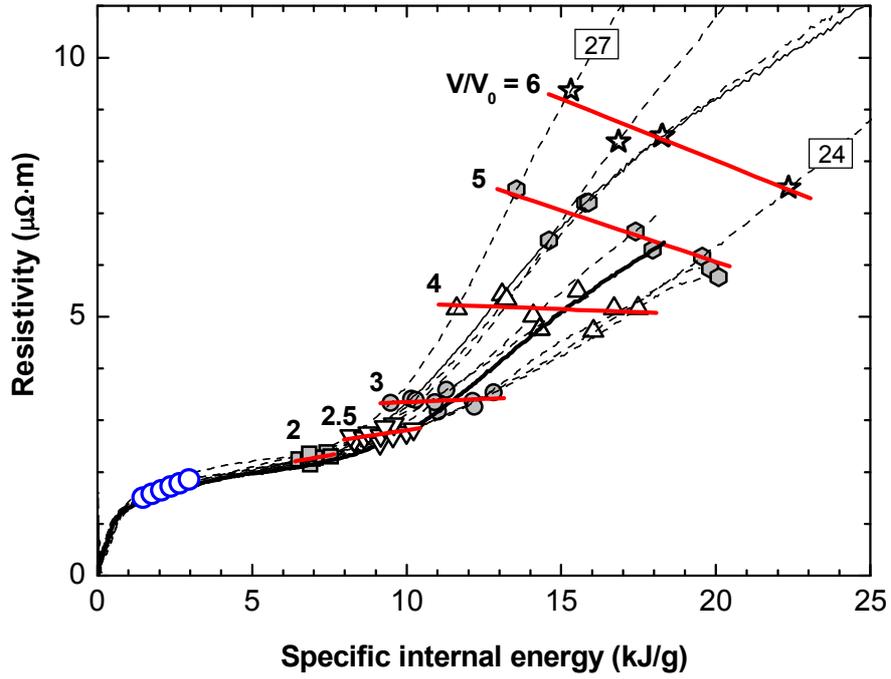

**Fig. 2.** (Color online) Resistivity of iron as a function of specific internal energy for 6 fixed values of the relative volume $V/V_0$. Our measurement results for the isochores are shown with black (gray and open) marks, red lines are linear fits to these data (the numbers are the values of relative volume for the isochores), dashed lines are dependencies measured in our experiments and blue open circles represent data [9]. Thick black line indicates an experiment in which the maximum pressure reached was 5 GPa.

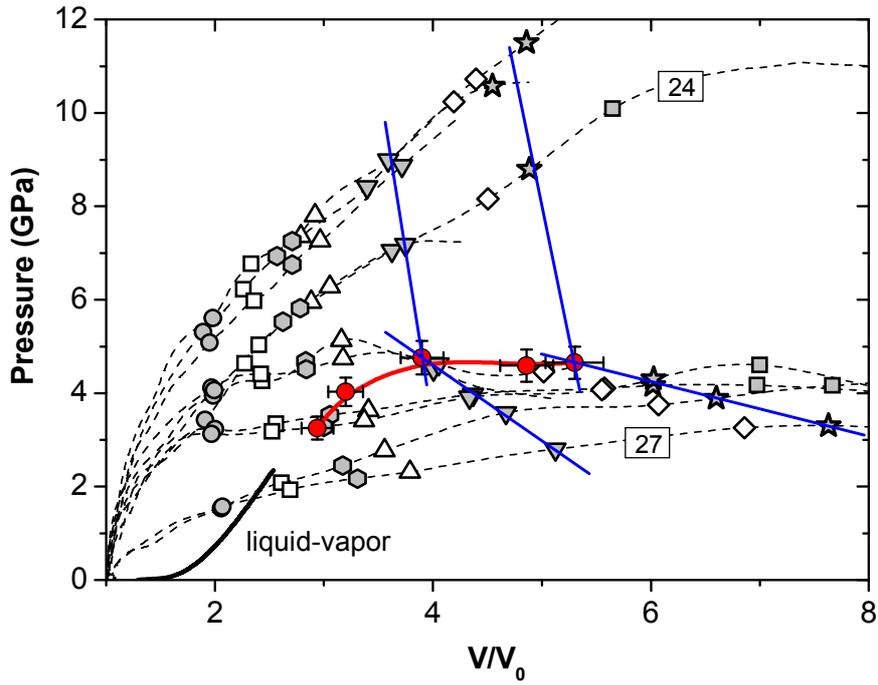

**Fig. 3.** (Color online) Pressure as a function of relative volume for 8 fixed values of entropy. The dashed lines represent dependencies obtained in our experiments, black (gray and open) marks correspond to the fixed entropy values, red marks indicate the points where the isentropes have kinks, the thick red line shows a smooth fit to these points. The blue straight lines show linear fits for two isentropes to determine the kinks positions. The thick black line shows a portion of the liquid-vapor equilibrium line which merges smoothly with that obtained here.

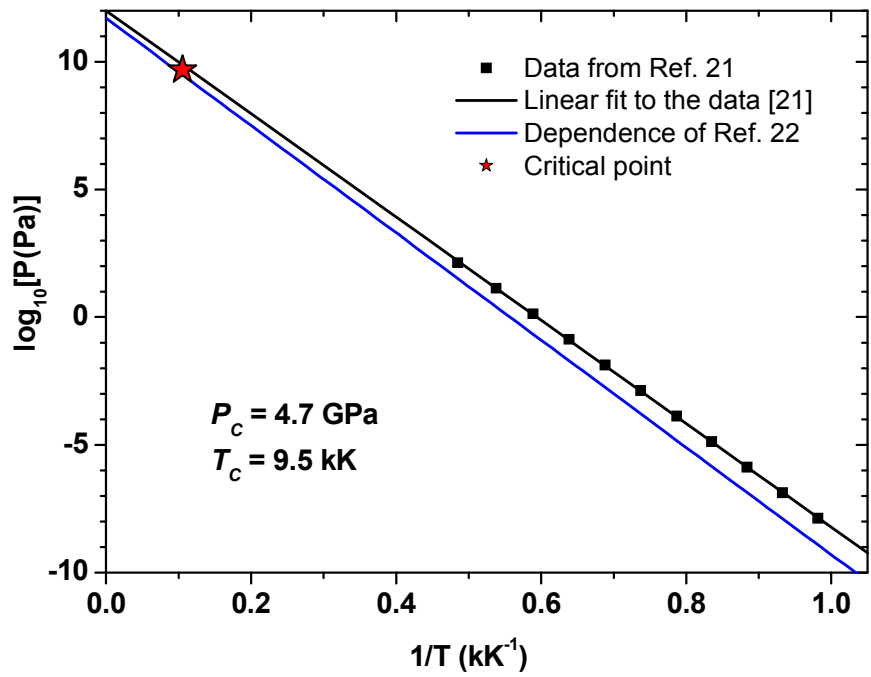

**Fig. 4.** (Color online)The saturated vapor pressure data of iron [21, 22] and the measured here critical pressure of 4.7 GPa are used to estimate the critical temperature.